\documentclass[manuscript]{aastex6}
\usepackage{longtable, enumerate, graphics, bm, graphicx}

\begin{document}

\title{On the Energetics of the $\mathrm{HCO}^+ + \mathrm{C} \rightarrow
  \mathrm{CH}^+ + \mathrm{CO}$ Reaction and Some Astrochemical Implications}

\author{D.W. Savin\altaffilmark{1}, R.G. Bhaskar\altaffilmark{1},
  S. Vissapragada\altaffilmark{1} and X. Urbain\altaffilmark{2}}
\altaffiltext{1}{\small Columbia Astrophysics Laboratory, Columbia
  University, New York, NY 10027, U.S.A.}
\altaffiltext{2}{\small Institute of Condensed Matter and
  Nanosciences, Universit\'e catholique de Louvain, B-1348
  Louvain-la-Neuve, Belgium}
\email{savin@astro.columbia.edu}

\begin{abstract}

  We explore the energetics of the titular reaction, which current
  astrochemical databases consider open at typical dense molecular
  (i.e., dark) cloud conditions.  As is common for reactions involving
  the transfer of light particles, we assume that there are no
  intersystem crossings of the potential energy surfaces involved.  In
  the absence of any such crossings, we find that this reaction is
  endoergic and will be suppressed at dark cloud temperatures.
  Updating accordingly a generic astrochemical model for dark clouds
  changes the predicted gas-phase abundances of 224 species by greater
  than a factor of 2.  Of these species, 43 have been observed in the
  interstellar medium.  Our findings demonstrate the astrochemical
  importance of determining the role of intersystem crossings, if
  any, in the titular reaction.


\end{abstract}



\section{Introduction}

The chemistry of molecular clouds involves a complicated interplay of
gas-phase reactions, chemistry on bare dust-grain surfaces, and
processes on and in the icy mantles of grains
\citep{Herbst:2014:PCCP,vanDishoeck:2014:FD}. Accurate chemical data
are necessary for each of these domains.  Experimental and theoretical
work can provide rate coefficients for gas-phase chemistry.  The
situation is more challenging for grain-surface or ice chemistry,
where the greater complexity and number of unknowns limit our ability
to generate the needed rate coefficient data. Hence, it is critical to
understand gas-phase reactions, in part so as to determine the
importance and extent of grain-surface and ice chemistry. By comparing
abundances from gas-phase astrochemical models to observations, one
can determine whether or not the inferred abundance for a given
molecule can be explained solely by gas-phase chemistry; if not, then
then that implies that either dust or ices are important or that the
gas-phase chemical data are inaccurate.

Gas-phase chemistry of clouds in the cold interstellar medium (ISM) is
simplified by the typical densities and temperatures
\citep{Snow:2006:ARAA,Herbst:2008:Book}.  Due to the low densities,
$\sim 10^3-10^7$~cm$^{-3}$, only two-body processes are important.
The low temperatures, $\sim 10-100$~K, rule out many neutral-neutral
chemical reactions, as they generally possess significant activation
energies. Hence, ion-neutral reactions (which typically possess no
activation barrier) are extremely important; almost two-thirds of all
reactions in current gas-phase astrochemical models are ion-neutral
reactions \citep[e.g.,][]{McElroy:2013:AA,Wakelam:2015:ApJS}.

In dark clouds, gas-phase chemistry is initiated by cosmic ray
ionization of H$_2$ \citep{Herbst:1973:ApJ,Watson:1973:ApJL}.  Nearly
all the resulting H$_2^+$ go on to react exoergically with another
H$_2$ molecule.  This occurs on a time scale that is quite rapid
compared to the life-time of a molecular cloud.  The resulting H$_3^+$
drives much of the ion-neutral chemistry in the cloud.
Photoionization of H$_2$ is unimportant as ultraviolet and soft X-rays
are strongly attenuated by the H and H$_2$ in the outer layers of the
cloud; additionally, the photoionization cross sections due to hard
X-rays and $\gamma$-rays are sufficiently small so as to minimize
their contribution \citep{Oka:2013:ChRv}.

H$_3^+$ can then readily react with CO, the second most abundant
neutral molecule in dark clouds \citep{Garrod:2007:AA}, via
\begin{equation}
    \mathrm{H_3}^+ + \mathrm{CO} \rightarrow \mathrm{HCO}^+ + \mathrm{H_2}.
\end{equation}
For ``typical'' dark cloud conditions (defined below), this reaction
is important for cloud ages $\gtrsim~10^5$~yrs.  The resulting HCO$^+$
is the most abundant molecular ion in dark clouds
\citep{Agundez:2013:ChRv}.  It has been detected in many such clouds,
including TMC-1(CP) and L134N, which have estimated ages of $\sim
10^5$~yrs \citep{Garrod:2007:AA}.

The role of HCO$^+$ in the ISM has recently been briefly reviewed by
\citet{Hamberg:2014:JPhChA}.  The ion readily transfers its proton to
many neutral molecules, thereby affecting the chemistry of molecular
clouds.  HCO$^+$ is also the dominant carrier of positive charge in
dark clouds and is used to probe the degree of ionization of the cloud
\citep{Agundez:2013:ChRv}.  This is important as the dynamics of the
cloud are modified by the presence of charge, which can couple to any
ambient magnetic field, affecting the transfer of angular momentum and
the dissipation of turbulence \citep{Dalgarno:2006:PNAS}.  The
strength of this coupling is determined by the fractional ionization
of the cloud.  Hence, our knowledge of dark clouds and their evolution
hinges, in part, on an accurate understanding of the underlying
chemistry controlling the HCO$^+$ abundance.

To that end, we have investigated the gas-phase astrochemistry of
HCO$^+$ using the Nahoon code \citep{Wakelam:2012:ApJS} combined with
the KInetic Database for Astrochemistry
\citep[KIDA;][]{Wakelam:2015:ApJS}.  Our initial studies indicated
that for dark cloud ages of $\sim 10^5$~yrs, the two most important
HCO$^+$ destruction mechanisms are dissociative recombination (DR)
with electrons via
\begin{equation}
  \label{eq:HCO++e-}
  \mathrm{HCO}^+ + \mathrm{e}^- \rightarrow \mathrm{neutral\ products},
\end{equation}
and the ion-neutral reaction with atomic C
\begin{equation}
  \label{eq:HCO++C}
  \mathrm{HCO}^+ + \mathrm{C} \rightarrow \mathrm{CH}^+ + \mathrm{CO}.
\end{equation}
Reaction~(\ref{eq:HCO++e-}), DR of HCO$^+$, has been extensively
studied both theoretically and experimentally \citep[for a review
  see][]{Hamberg:2014:JPhChA}.  Though some issues remain, it is
thought to be relatively well understood.  The same cannot be said for
Reaction~(\ref{eq:HCO++C}), for which there appear to be no
theoretical or experimental studies.  Present-day astrochemical models
use the recommended rate coefficient of \citet{Prasad:1980:ApJS},
which appears to be an estimate based on the Langevin formalism.

Given the apparent importance of Reaction~(\ref{eq:HCO++C}), we
explored the possibility of measuring it in our laboratory.  Recently
we have developed a novel dual-source, merged-beams apparatus for
studying ion-neutral reactions.  With this device we have measured
reactions of H$_3^+$ with atomic C and O
\citep{OConnor:2015:ApJS,deRuette:2016:ApJ} and investigated the
astrochemical implications of our new chemical data
\citep{deRuette:2016:ApJ,Vissapragada:2016:ApJ}.  So it seemed a
natural extension of that work to study Reaction~(\ref{eq:HCO++C}).
However, as we investigated the energetics of this reaction, we
quickly realized that it was unlikely to be exoergic; rather, it is
more likely to endoergic by an amount sufficiently large for this
channel to be closed at molecular cloud temperatures.  This raises the
question: what are the astrochemical implications of
Reaction~(\ref{eq:HCO++C}) being endoergic?

In the rest of this paper we explore these implications.
Section~\ref{sec:Energetics} reviews the energetics of
Reaction~(\ref{eq:HCO++C}).  Section~\ref{sec:Modeling} briefly
discusses our dark cloud astrochemical model.
Section~\ref{sec:Implications} presents the results of our modeling
and discusses some of the astrochemical implications of
Reaction~(\ref{eq:HCO++C}) being closed at dark cloud temperatures.
Lastly, Section~\ref{sec:Summary} summarizes our findings.

\section{Energetics}
\label{sec:Energetics}

In the cold ISM, molecules reside primarily in their lowest electronic
state (X) and lowest vibrational level(s).  The rotational populations
are more sensitive to the temperature and density of the gas, but the
bulk of the population typically resides in the lowest rotational
levels.  Rotational excitations have a negligible effect on the
energetics of Reaction~(\ref{eq:HCO++C}).  For our calculations here,
we assume all parent and daughter molecules are in their lowest
electronic, vibrational, and rotational levels.

The neutral C in dark clouds is of $^3$P symmetry.  The HCO$^+$ has a
X~$^1\Sigma^+$ symmetry.  Taking into account spin multiplicities
\citep{Talbi:1991:ApJ} and making the common assumption for reactions
involving the transfer of light particles that intersystem crossings
of the potential energy surfaces do not lead to substantial
redistribution of the flux among the different spin symmetries
\citep{Salem:1972:ACIE,Li:2014:JPhChA,Martinez:2015:JChPh}, the two
lowest energy channels for Reaction~(\ref{eq:HCO++C}) at 0~K are
\begin{eqnarray}
  \label{eq:lowestE}
  \mathrm{HCO}^+(\mathrm{X}\ ^1\Sigma^+) + \mathrm{C}(^3\mathrm{P})
  & \rightarrow &
  \mathrm{CH}^+(\mathrm{a}\ ^3\Pi) + \mathrm{CO}(\mathrm{X}\ ^1\Sigma^+)
  + \Delta E, \\
  \label{eq:nextlowestE}
    & \rightarrow &
  \mathrm{CH}^+(\mathrm{X}\ ^1\Sigma^+) + \mathrm{CO}(\mathrm{a}\ ^3\Pi)
  + \Delta E.
\end{eqnarray}
Here $\Delta E$ is the reaction energy, defined here as the change in
the total internal energies of the reactants and daughter products.
$\Delta E$ is negative for endoergic reactions and positive for
exoergic reactions.

The energies needed to calculate the energetics for
Reactions~(\ref{eq:lowestE}) and (\ref{eq:nextlowestE}) are given in
Table~\ref{tab:energetics}.  The dissociation energy from the
molecular potential minimum is $D_{\rm e}$.  For HCO$^+$, we take the
value from \citet{Mladenovic:1998:JChPh} for the case where the H$^+$
is bound to the C.  For CH$^+$, we use the data of
\citet{Barinovs:2004:CPL} and for CO that of \citet{Shi:2013:IJQC}.
For the reaction energetics, the quantity needed is the dissociation
energy from the lowest ro-vibrational level, $D_0$, which is $D_{\rm
  e}$ minus the zero-point energy (ZPE).
\citet{Hechtfischer:2002:JChPh} measured $D_0$ for
CH$^+$(X\ $^1\Sigma^+$).  We cal-

\begin{deluxetable}{ccccc}
\label{tab:energetics}
\tablecaption{Quantities Needed to Determine the Energetics for
  Reactions~(\ref{eq:lowestE}), (\ref{eq:nextlowestE}), and
  (\ref{eq:intersystem}).}
\tablewidth{0pt}
\tablehead{
\colhead{} &
\colhead{$D_e$} &
\colhead{ZPE} &
\colhead{$D_0$} &
\colhead{$|\Delta E|$} \\
\colhead{Process} & \colhead{(eV)} & \colhead{(eV)} & \colhead{(eV)} & \colhead{(eV)}
}
\tablecolumns{5}
\startdata
HCO$^+$(X$^1\Sigma^+$) $\rightarrow$
    {H}$^+$($^1$S) + CO(X$^1\Sigma^+$) &
    6.400 & 0.437 & 5.963 & \nodata \\
CH$^+$(X$^1\Sigma^+$) $\rightarrow$
    {C}$^+$($^2$P) + {H}($^2$S) &
    4.264 & \nodata & 4.085 & \nodata\\
CH$^+$(a$^3\Pi$) $\rightarrow$
    {C}$^+$($^2$P) + {H}($^2$S)
    & 3.068 & 0.167 & 2.901 & \nodata\\
CO(X$^1\Sigma^+$) $\rightarrow$
    {C}($^3$P) + {O}($^3$P) &
    11.224 & 0.134 & 11.090 & \nodata\\
CO(a$^3\Pi$) $\rightarrow$
    {C}($^3$P) + {O}($^3$P) &
    5.188 & 0.109 & 5.079 & \nodata\\
CO(X$^1\Sigma^+$) $\rightarrow$
    CO(a$^3\Pi$) &
    \nodata & \nodata & \nodata & 6.011 \\
C$^+$($^2$P) + {H}($^2$S) $\rightarrow$
    {C}($^3$P) + H$^+$($^1$S) &
    \nodata & \nodata & \nodata & 2.338
\enddata
\tablecomments{$D_{\rm e}$ is the dissociation energy from the
  molecular potential minimum, ZPE is the zero-point energy of the
  lowest ro-vibrational level of the molecule, $D_0$ is the
  dissociation energy from this lowest level, and $\Delta E$ is the
  reaction energy.  The various energies are taken from
  \citet{Huber:1976:NIST},
  \citet{Mladenovic:1998:JChPh},
  \citet{Barinovs:2004:CPL},
  \citet{Hechtfischer:2002:JChPh},
  \citet{Irikura:2007:JPCRD},
  \citet{Shi:2013:IJQC},
  \citet{Mladenovic:2014:AA},
  \citet{NIST_ASD}, and
  M.\ Delsaut \& J. Li\'evin (in preparation).}
\end{deluxetable}

\noindent culate $D_0$ for the other systems using the ZPE for HCO$^+$ from
\citet{Mladenovic:2014:AA}, for CH$^+$(a$^3\Pi$) from M.\ Delsaut \&
J. Li\'evin (in preparation), and for CO(X\ $^1\Sigma^+$) from
\citet{Irikura:2007:JPCRD}.  For CO(a\ $^3\Pi$), we have calculated
the ZPE using the molecular constants of \citet{Huber:1976:NIST}.  We
also note that the CH$^+$ $D_0$ is for dissociation to
$\mathrm{C^+(^2P) + H(^2S)}$.  Reactions~(\ref{eq:lowestE}) and
(\ref{eq:nextlowestE}) involve $\mathrm{H^+(^1S)}$ bonding to the
$\mathrm{C(^3P)}$.  Using the energies from \citet{NIST_ASD}, these
lie an additional 2.388~eV above the products of the CH$^+$
dissociation limit.

From the information above, we can readily calculate $\Delta E$ for
Reactions~(\ref{eq:lowestE}) and (\ref{eq:nextlowestE}).  For
Reaction~(\ref{eq:lowestE}) we loose energy dissociating
$\mathrm{HCO^+(X^1\Sigma^+)}$ and gain energy going from
$\mathrm{C(^3P)+H^+(^1S)}$ to $\mathrm{CH^+(a^3\Pi)}$.  This gives
$\Delta E = -5.963~{\rm eV} + 2.901~{\rm eV} + 2.338~{\rm eV} =
-0.724~{\rm eV}$, where the negative energy means that the reaction is
endoergic.  In a more chemical notation, the enthalpy of the reaction
at 0~K is $\Delta_{\rm r} H_0 = 69.9$~kJ/mol, where a positive
enthalpy signifies that the reaction is endoergic.  The calculations
are similar for Reaction~(\ref{eq:nextlowestE}), except now we gain
energy forming $\mathrm{CH^+(X^1\Sigma^+)}$ and loose energy exciting
the CO to the $\mathrm{a^3\Pi}$ symmetry.  This gives $\Delta E =
-5.963~{\rm eV} + 4.086~{\rm eV} + 2.338~{\rm eV} -6.011~{\rm eV} =
-5.550~{\rm eV}$.  The corresponding enthalpy of the reaction at 0~K
is $\Delta_{\rm r} H_0 = 535.4$~kJ/mol.  Hence both reactions are
endoergic by an amount $\Delta E/k_{\rm B}=-8,402$~K for
Reaction~(\ref{eq:lowestE}) and $-64,405$~K for
Reaction~(\ref{eq:nextlowestE}), where $k_{\rm B}$ is the Boltzmann
constant.  All these quantities have been calculated for 0~K.  The
energetics at dark cloud temperatures of $\sim 10$~K are essentially
the same.

From this analysis, it is not clear why \citet{Prasad:1980:ApJS} treat
Reaction~(\ref{eq:HCO++C}) as being open with a thermal rate
coefficient of $\mathrm{1.1 \times 10^{-9}~cm^3~s^{-1}}$.  In their
compilation of reactions for gas phase chemistry in interstellar
clouds, they do not discuss the issues of spin multiplicities or
intersystem crossings.  It seems likely that they did not consider
these issues but rather just assumed that the CH$^+$ and CO both
formed in their ground symmetries, namely
\begin{equation}
  \label{eq:intersystem}
  \mathrm{HCO}^+(\mathrm{X}^1\Sigma^+) + \mathrm{C}(^3\mathrm{P})
  \rightarrow 
  \mathrm{CH}^+(\mathrm{X}^1\Sigma^+) + \mathrm{CO}(\mathrm{X}^1\Sigma^+)
  + \Delta E. 
\end{equation}
Here, the reaction would be exoergic by $\Delta E = -5.963~{\rm eV} +
4.085~{\rm eV} + 2.338~{\rm eV} = 0.460~{\rm eV}$ or $\Delta E/k_{\rm
  B}=5,338$~K.  The enthalpy would be $\Delta_{\rm r} H_0 =
-44.4$~kJ/mol, where the negative sign signifies that the reaction is
exoergic.  But this channel would be open only if an intersystem
transition occured during the reaction.
For now we follow the common practice of assuming that intersystem
crossings are not important for reactions involving the transfer of
light particles.  Thus, below we assume that
Reaction~(\ref{eq:HCO++C}) does not proceed in the cold ISM.
Definitively resolving this issue will likely require detailed
theoretical and experimental chemical studies.

Lastly, we note that the enthalpy for some of the above reactions can
also be calculated using the Active Thermochemical Tables hosted at
Argonne National Laboratory\footnote{http://atct.anl.gov/}.  Those
table do not provide the needed data for Reaction~(\ref{eq:lowestE});
but they do give $\Delta_r H_0$ for Reactions~(\ref{eq:nextlowestE})
and (\ref{eq:intersystem}) as $546.13 \pm 0.10$~kJ/mol and $-33.25 \pm
0.01$~kJ/mol, respectively.  These are about 11~kJ/mol larger than our
values derived here, but do not change any of our conclusions about
the energetics.

\section{Dark Cloud Model}
\label{sec:Modeling}

We adopt here the generic dark cloud conditions given by
\citet{Wakelam:2015:ApJS}, using their initial chemical abundances, a
visual extinction of $A_{\rm v} = 30$, a hydrogen nuclei number
density of $n_{\rm H} = 10^4$~cm$^{-3}$, a temperature of 10~K, and a
cosmic ray ionization rate for H$_2$ of $\zeta = 10^{-17}$~s$^{-1}$.
The chemical evolution of the cloud is calculated using Nahoon
\citep{Wakelam:2012:ApJS} and a version of KIDA
\citep{Wakelam:2015:ApJS} which we have updated as described below.
KIDA includes 489 species and over 7500 reactions.

We have modified KIDA slightly by incorporating our
experimentally-derived thermal rate coefficient results from
\citet{OConnor:2015:ApJS} for the reactions
\begin{eqnarray}
\mathrm{C} + \mathrm{H_3}^+ & \rightarrow & \mathrm{CH}^+ + \mathrm{H_2},\\
                            & \rightarrow & \mathrm{CH_2}^+ + \mathrm{H}.
\end{eqnarray}
Additionally, we use our experimental results from \citet{deRuette:2016:ApJ}
for
\begin{eqnarray}
\mathrm{O} + \mathrm{H_3}^+ & \rightarrow & \mathrm{OH}^+ + \mathrm{H_2},\\
                            & \rightarrow & \mathrm{H_2O}^+ + \mathrm{H}.
\end{eqnarray}
However, in this case we only extracted the thermal rate coefficient
for the sum of both channels.  Here we have assumed branching ratios
of 100\%:0\% and 0\%:100\% for forming OH$^+$:H$_2$O$^+$ and find no
difference in the results of our astrochemical simulations.  We
attribute this to OH$^+$ and H$_2$O$^+$ both undergoing rapid
sequential hydrogen abstraction with the abundant H$_2$ to form
H$_3$O$^+$.

\section{Astrochemical Implications}
\label{sec:Implications}

\begin{figure}
    \centering
    \includegraphics[width=1\textwidth]{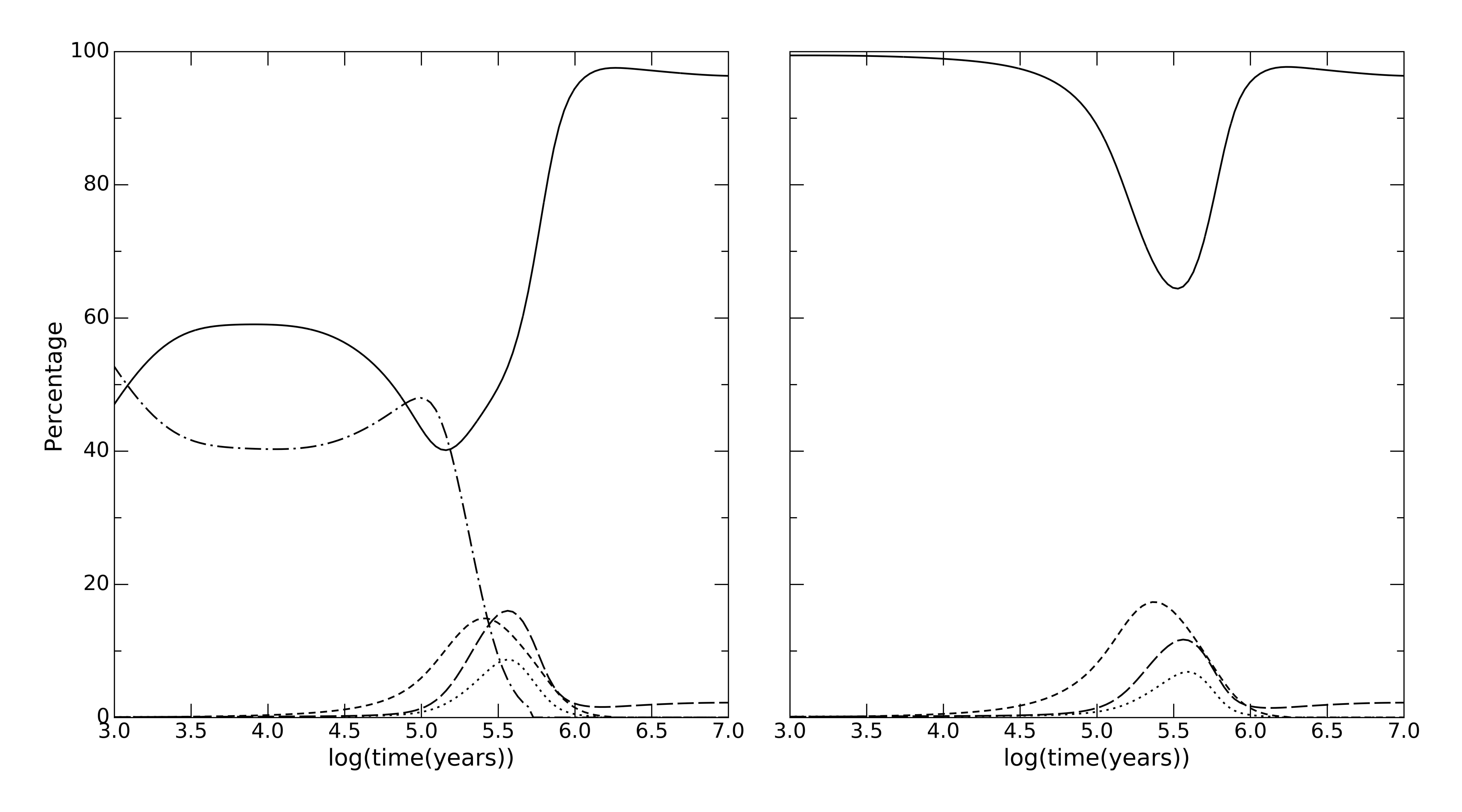}
    \caption{Most important HCO$^+$ destruction mechanisms, in
      percentage, for a generic dark cloud as afunction of cloud age.
      The left panel shows the destruction mechanisms when
      Reaction~(\ref{eq:HCO++C}) is treated as open and the right
      panel when the reaction is closed.  The solid curve is for DR
      (Reaction~\ref{eq:HCO++e-}), the dot-dashed curve is for the
      reaction with atomic C (Reaction~\ref{eq:HCO++C}), the
      short-dashed curve is for Reaction~(\ref{eq:HCO++C3}), the
      long-dashed curve is for Reaction~(\ref{eq:HCO++H2O}), and the
      dotted curve is for Reaction~(\ref{eq:HCO++HCN}).}
    \label{fig:HCO+Destruction}
\end{figure}

Figure~\ref{fig:HCO+Destruction} shows the dominant HCO$^+$
destruction mechanisms for our generic dark cloud as a function of the
cloud age.  The left panel shows the destruction mechanisms when
Reaction~(\ref{eq:HCO++C}) is treated as open and the right panel when
the reaction is closed.  If the reaction were indeed open, then the
two dominant destruction mechanisms would be
Reactions~(\ref{eq:HCO++e-}) and (\ref{eq:HCO++C}), with
Reaction~(\ref{eq:HCO++C}) becoming unimportant after $10^{5.3}$~yr as
the free atomic C becomes bound up into molecules.  However, our
energetics study indicates that Reaction~(\ref{eq:HCO++C}) is closed,
and that DR is the dominant HCO$^+$ destruction mechanism for all
cloud ages.

In either case, for cloud ages between $\sim 10^5-10^6$~yr, HCO$^+$
can also be destroyed via the minor reactions
\begin{eqnarray}
  \label{eq:HCO++C3}
  \mathrm{HCO}^+ + \mathrm{C_3}
  & \rightarrow & \mathrm{CO} + \mathrm{C_3H}^+, \\
  \label{eq:HCO++H2O}
  \mathrm{HCO}^+ + \mathrm{H_2O}
  & \rightarrow & \mathrm{CO} + \mathrm{H_3O}^+,
\end{eqnarray}
and
\begin{eqnarray}
  \label{eq:HCO++HCN}
  \mathrm{HCO}^+ + \mathrm{HCN}
  & \rightarrow & \mathrm{CO} + \mathrm{HCNH}^+.
\end{eqnarray}
The percentage contribution of these reactions to the total HCO$^+$
destruction rate are also shown in Figure~\ref{fig:HCO+Destruction}.

\begin{figure}
    \centering
    \includegraphics[width=1\textwidth]{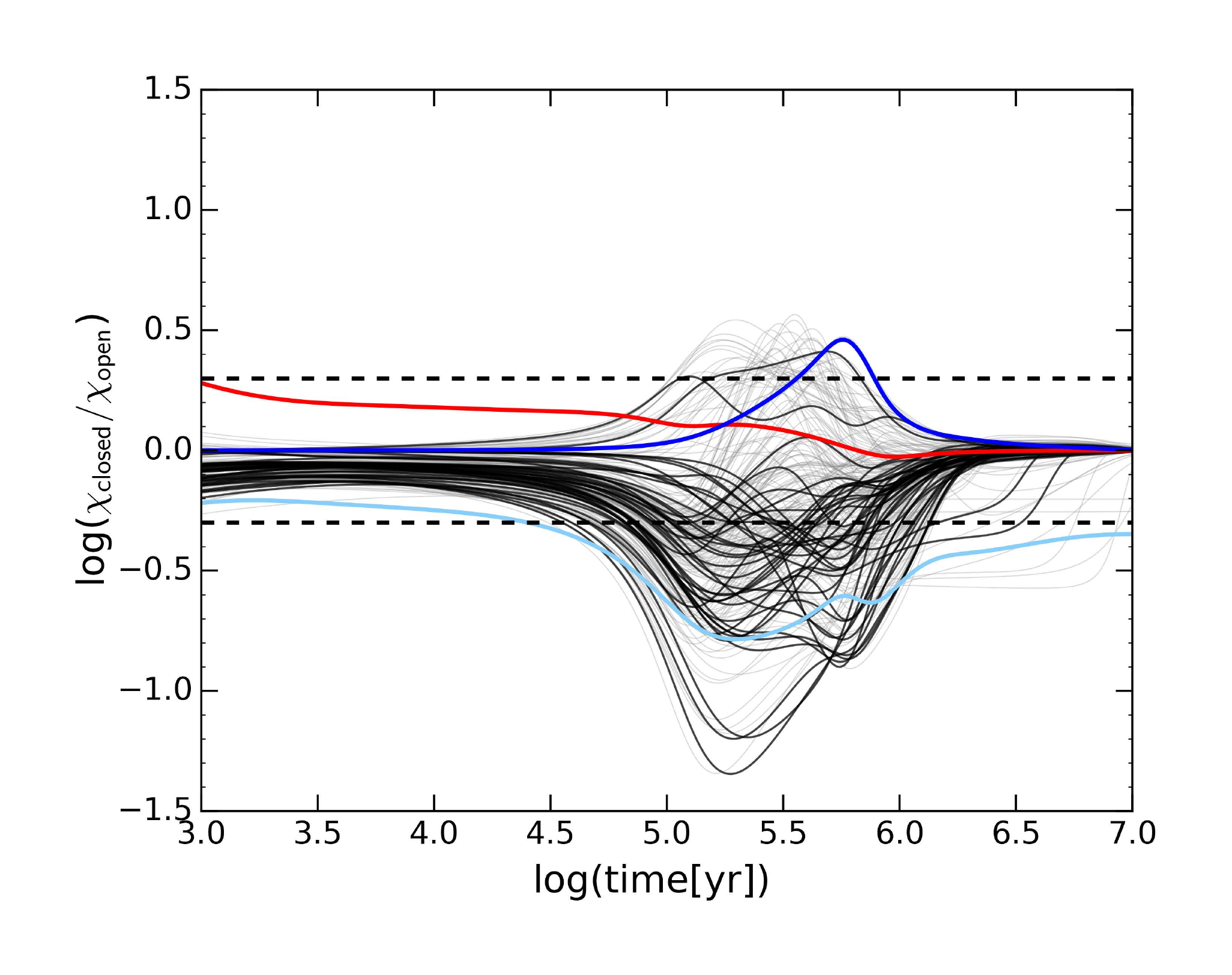}
    \caption{Ratios of the predicted abundances for all species in
      KIDA that are significantly impacted by the closing of
      Reaction~(\ref{eq:HCO++C}) for cloud ages between $10^5$ to
      $10^6$~yr.  Abundances were calculated for the channel being
      closed ($\chi_{\rm closed}$) relative to those using the
      unmodified KIDA database which assumes that the channel is open
      ($\chi_{\rm open}$).  The log of the ratio is plotted as a
      function of cloud age for the initial conditions discussed in
      the text.  The dashed lines represent the thresholds for
      ``significantly'' affected species (those whose abundance
      changes by a factor of 2 or more).  The red curve shows the
      abundance ratio for HCO$^+$, the dark blue for C, and the
      light blue for CH$^+$.  The remaining gray and black curves
      are for all of the 222 other significantly affected species in
      KIDA, with the black curves showing the 43 of these
      that have been observed in the ISM.}
    \label{fig:ratios}
\end{figure}

We have also calculated the predicted abundances for all 489 species
in KIDA.  The abundances were calculated with
Reaction~(\ref{eq:HCO++C}) closed ($\chi_{\rm closed}$) and open
($\chi_{\rm open}$).  The latter is currently assumed by astrochemical
databases.  Figure~\ref{fig:ratios} shows the $\log$ of the relative
abundance ratios $\chi_{\rm closed}/\chi_{\rm open}$, where we plot
only those species whose abundance ratio changes by a factor of 2 or
more for cloud ages between $10^5$ to $10^6$~yr.

The structure seen in Figure~\ref{fig:ratios} can readily be explained
by the abundances changes of CH$^+$, C, and HCO$^+$.  The abudance of
CH$^+$ is reduced by the closing of Reaction~(\ref{eq:HCO++C}) As a
result, the predicted abundances increase for those
species which are destroyed by reactions with CH$^+$.  Closing
Reaction~(\ref{eq:HCO++C}) also generally increases the abundances of
C and HCO$^+$; and the predicted abundances increase for those species
which are formed through reactions involving C and/or HCO$^+$.
Conversely, the abundances decrease for those species requiring CH$^+$
to form and/or if the precursors to these species are destroyed in
reactions with C and/or HCO$^+$.

For typical observed cloud ages of between $10^5$
to $10^6$~yr, we find that the predicted abundances of 224 species
change by more than a factor of 2.  Of these species, 43 have been
observed in the ISM.  Many of these are predicted to form in the gas
phase \citep{Walsh:2009:ApJ,Agundez:2013:ChRv} such as
the neutral hydrocarbons CH$_3$, CH$_4$, C$_2$H$_2$, C$_2$H$_4$ and CH$_3$CCH;
the amines CH$_2$NH$_2$ and CH$_3$NH$_2$;
the cyanides and isocyanides HNC, CH$_3$CN, H$_2$CCN and HNC$_3$;
the polyynes and methylpolyynes C$_4$H$_2$, CH$_3$C$_4$H and CH$_3$C$_6$H;
the cyanopolyynes HC$_4$N, HC$_5$N and CH$_3$C$_3$N; and
the molecular cations CH$^+$, HCNH$^+$, H$_2$COH$^+$ and HC$_3$NH$^+$.

\section{Summary}
\label{sec:Summary}

HCO$^+$ is an important ion in the chemical and physical evolution of
dark molecular clouds.  We have explored the energetics of
Reaction~(\ref{eq:HCO++C}), which has long been assumed to be exoergic
\citep{Prasad:1980:ApJS} and as a result also appeared to be
astrochemically important.  However, in reactions involving the
transfer of light particles it is commonly assumed that intersystem
crossings are unimportant.  If that is the case, then our results
indicate that Reaction~(\ref{eq:HCO++C}) is endoergic and will not
proceed at typical molecular cloud temperatures.  Our modeling of a
generic dark cloud with this channel closed indicates that DR is the
dominant destruction mechanism of HCO$^+$ at all cloud ages.  We also
find that the predicted abundances of 224 species change by greater
than a factor of 2 as a result of closing Reaction~(\ref{eq:HCO++C}).
Our findings demonstrate the astrochemical importance for determing
the role of intersystem crossings in Reaction~(\ref{eq:HCO++C}).

As a final point, our findings are unlikely to have any impact on the
long-standing issue of HCO$^+$ and CH$^+$ abundances in diffuse clouds
\citep{Godard:2010:AA,Valdivia:2017:AA}.  The observed abunances are
one to two orders of magnitude larger than predicted by UV-dominated
chemical models that include Reaction~(\ref{eq:HCO++C}).  If this
reaction were closed, that would reduce the HCO$^+$ destruction rate,
thereby increasing the predicted abundance.  However, this is unlikely
to result in more than a factor of a couple increase as DR is more
likely to be the dominant HCO$^+$ destruction mechanism in either
case.  As for CH$^+$, if Reaction~(\ref{eq:HCO++C}) were closed, that
would decrease the predicted CH$^+$ abundance, thereby further
increasing the existing discepancy between obsevations and models.

\acknowledgements

The authors thank K.\ Bowen, P.-M.\ Hillenbrand, J.\ Li\'evin,
D. C. Lis, O.\ Novotn\'y, I.\ R.\ Sims, and A.\ Viggiano for
stimulating discussions.  D.W.S.\ and R.G.B.\ were supported in part
by the NASA Astrophysics Research and Analysis (APRA) program and the
NSF Division of Astronomical Sciences Astronomy and Astrophysics
Grants (AAG) program.  S.V.\ was supported in part by a Barry
Goldwater Scholarship and the USRA James B. Willett Educational
Memorial Scholarship Award.  X.U.\ is a Senior Research Associate of
the FRS-FNRS.

\bibliographystyle{aasjournal}
\bibliography{Bibliography}

\end{document}